\documentclass[%
 aps,pre,
 amsmath,amssymb,
 reprint,%
]{revtex4-2}
\usepackage{pdfpages}
\usepackage{pgffor}
\makeatletter
\AtBeginDocument{\let\LS@rot\@undefined}
\makeatother

\usepackage{graphicx}
\usepackage{dcolumn}
\usepackage{bm}

\usepackage[utf8]{inputenc}
\usepackage[T1]{fontenc}
\usepackage{etoolbox}
\usepackage{mathtools}
\usepackage{xcolor}

\def\@email#1#2{%
 \endgroup
 \patchcmd{\titleblock@produce}
  {\frontmatter@RRAPformat}
  {\frontmatter@RRAPformat{\produce@RRAP{*#1\href{mailto:#2}{#2}}}\frontmatter@RRAPformat}
  {}{}
}%
\usepackage{wrapfig}
\usepackage{float}
\usepackage{physics}
\usepackage{bbold}
\usepackage{tabularx}
\usepackage[unicode=true,bookmarks=true,bookmarksnumbered=false,bookmarksopen=false, breaklinks=false,pdfborder={0 0 0},pdfborderstyle={},backref=false,colorlinks=true]{hyperref}

\hypersetup{pdfborderstyle={},pdfborderstyle={},pdfborderstyle={},pdfborderstyle={},pdfborderstyle={},pdfborderstyle={},linkcolor=blue,citecolor=blue}

\usepackage{mathpazo}

\newcommand{\Tb}{T_{\text{b}}}
\newcommand{\Tbm}{T_{\text{b}}^-}
\newcommand{\TbmA}{T_{\text{b,A}}^-}
\newcommand{\TbmB}{T_{\text{b,B}}^-}
\newcommand{\Tbp}{T_{\text{b}}^+}
\newcommand{\Tm}{T_{\text{b}}^{-*}}
\newcommand{\TmB}{T_{\text{b,B}}^{-*}}
\newcommand{\Tp}{T_{\text{b}}^-}
\newcommand{\TA}{T_{\text{A}}}
\newcommand{\TB}{T_{\text{B}}}
\newcommand{\Th}{\Tb^{\text{h}}}
\newcommand{\Tc}{\Tb^{\text{c}}}
\newcommand{\Tw}{\Tb^{\text{w}}}
\newcommand{\tw}{t_{\text{w}}}
\newcommand{\twth}{\tw^{\min}}
\newcommand{\tx}{t_{\times}}
\newcommand{\tM}{t_{\text{M}}}

\newcommand{\wT}{\widetilde{T}}
\newcommand{\wTb}{\widetilde{T}_{\text{b}}}
\newcommand{\wH}{\widetilde{H}}
\newcommand{\wEE}{\widetilde{\mathcal{E}}}

\newcommand{\EE}{\mathcal{E}}
\newcommand{\KK}{K}
\newcommand{\tauE}{$\tau$-exp }

\newcommand\beq{\begin{equation}}
\newcommand\eeq{\end{equation}}
\newcommand\beqa{\begin{eqnarray}}
\newcommand\eeqa{\end{eqnarray}}
\newcommand{\nn}{\nonumber\\}
\def\bal#1\eal{\begin{align}#1\end{align}}

\begin{document}

\title{Mpemba meets Newton: Exploring the Mpemba and Kovacs effects in the time-delayed cooling law
}
\author{Andr\'es Santos}%
 \affiliation{Departamento de F\'isica and Instituto de Computaci\'on Cient\'ifica Avanzada (ICCAEx), Universidad de Extremadura, E-06006 Badajoz, Spain}
\email{andres@unex.es}

\date{\today}

\begin{abstract}

Despite extensive research, the fundamental physical mechanisms underlying the Mpemba effect, a phenomenon where a substance cools faster after initially being heated, remain elusive. Although historically linked with water, the Mpemba effect manifests across diverse systems, sparking heightened interest in Mpemba-like phenomena.
Concurrently, the Kovacs effect, a memory phenomenon observed in materials such as polymers, involves rapid quenching and subsequent temperature changes, resulting in nonmonotonic relaxation behavior. This paper probes the intricacies of the Mpemba and Kovacs effects within the framework of the time-delayed Newton's law of cooling, recognized as a simplistic yet effective phenomenological model accommodating memory phenomena.
This law allows for a nuanced comprehension of temperature variations, introducing a delay time ($\tau$) and incorporating specific protocols for the thermal bath temperature, contingent on a defined waiting time ($\tw$). Remarkably, the relevant parameter space is two-dimensional ($\tau$ and $\tw$), with bath temperatures exerting no influence on the presence or absence of the Mpemba effect or the relative strength of the Kovacs effect.
 The findings enhance our understanding of these memory phenomena, providing valuable insights applicable to researchers across diverse fields, ranging from physics to materials science.
\end{abstract}

\maketitle
\section{Introduction}
As a teenager, Mpemba (1950--2023) accidentally discovered the paradoxical effect that now bears his name. In the first part of his renowned paper \cite{MO69} with Osborne (1932--2014), the 19-year-old Mpemba candidly recounted the story. In the second part of the paper, Osborne wrote,
\begin{quotation}
\noindent
``The headmaster of Mkwawa High School invited me to speak to the students on `Physics and national development'. [\ldots] One
student raised a laugh from his colleagues with a question I remember as `If you take two beakers with equal volumes of water, one at $35^\circ$C and the other at $100^\circ$C,  and put them into a refrigerator, the one that started at $100^\circ$C freezes first. Why?'.
It seemed an unlikely happening, but the student insisted that he was sure of the facts. I confess that I thought he was mistaken but fortunately remembered the need to encourage students to develop questioning and critical attitudes. No question should be ridiculed. In this case there was an added reason for caution, for everyday events are seldom as simple as they seem and it is dangerous to pass a superficial judgment on what can and cannot be. I said that the facts as they were given surprised me because they appeared to contradict the physics I knew. But I added that it was possible that the rate of cooling might be affected by some factor I had not considered. I promised I would put the claim to the test of experiment and encouraged my questioner to repeat the experiment himself.''
\end{quotation}
In that second part of Ref.~\cite{MO69}, Osborne reported some experimental results confirming the effect, although he conceded that ``The experiments attempted were relatively crude and several factors could influence cooling rates. More sophisticated experiments are needed to provide a more certain answer to the question.''

The above paragraph succinctly states what the Mpemba effect is about---the possibility that hot water freezes faster than cold water. But, more importantly, Osborne emphasizes the importance of fostering a spirit of curiosity and open-mindedness in scientific inquiry. When confronted with a seemingly counterintuitive claim by a student, Osborne resisted the temptation to dismiss or ridicule the idea outright. Instead, he  acknowledged the potential limitations of his own understanding and expressed a willingness to explore the claim through experimentation. This approach reflects a commitment to the principle that scientific inquiry should be driven by evidence and an openness to reevaluate established beliefs in the face of new and unexpected observations. Osborne's response underscores the idea that questioning assumptions and testing unconventional ideas can lead to a deeper understanding of the complexities inherent in scientific phenomena.

After 1969, attention to the Mpemba effect was mainly confined to popular science and education journals \cite{K69,F71,D71,F74,G74,W77,O79,F79,K80,H81,A95,K96,CK06,J06,K09,B11,WCVN11,BT12,S15,BT15,R15,IC16}. It was also revealed that the phenomenon had already been noted by classical philosophers and scholars, including Aristotle \cite{aristotle_works_1931}, Roger Bacon \cite{Bacon_Opus_Majus},  Francis Bacon \cite{Bacon1620}, and Descartes \cite{Descartes1637}. Interestingly, in the same year that Mpemba and Osborne published their paper, Kell independently published a brief note that began with the following statement: ``It is widely believed, at least in Canada, that hot water will freeze more quickly than cold water.''

A unanimous agreement remains elusive regarding the fundamental physical mechanisms responsible for the Mpemba effect. Various factors, including water evaporation \cite{K69,F71,W77,VM10}, natural convection \cite{D71,M96,IC16}, disparities in the gas composition of water \cite{F79,WOB88,K09},  the effects of supercooling, either independently \cite{A95,ZHMZZZJS14,SHZMW23} or in combination with other factors \cite{ERS08,VM12,VK15,JG15}, lack of energy equipartition \cite{GLH19}, or the influence of rough walls in creating nucleation sites \cite{BH20}, have been proposed as potential contributors to the Mpemba effect. Conversely, doubts have been raised about the very existence of the Mpemba effect in water \cite{BL16,BH20,ES21}.

\begin{figure}
      \includegraphics[width=0.9\columnwidth]{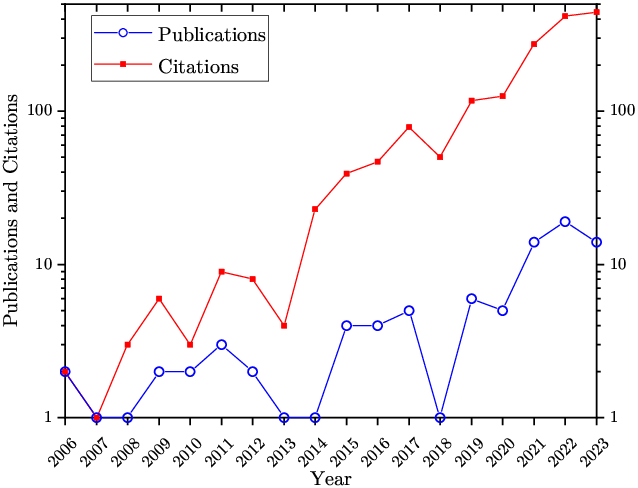}
      \caption{Number of publications and citations in the period 2006--2023 obtained from the Web of Science \cite{WoS24} with the search query ``topic=(Mpemba effect).''  Note the vertical logarithmic scale.
  \label{fig00}}
\end{figure}

In addition to its historical association with water, the Mpemba effect exhibits similar phenomena reported across a diverse range of systems, including
carbon nanotube resonators~\cite{GLCG11},
clathrate hydrates~\cite{AKKL16},
Markovian models~\cite{LR17,KRHV19,CKB21,BGM21,LLHRW22,TYR23b},
granular gases~\cite{LVPS17,TLLVPS19,BPRR20,MLTVL21,GKG21,GG21,BPR21,BPR22,MS22,MS22b,PSPP23},
molecular gases under drag~\cite{SP20,PSP21,MSP22,PSPP23},
spin glasses~\cite{Betal19},
non-Markovian mean-field systems~\cite{YH20,YH22},
inertial suspensions~\cite{THS21,T21},
colloidal systems~\cite{KB20,BKC21,CKB21,KCB22,IDLGR24},
Ising-like models~\cite{GMLMS21,VD21,TYR23a,TYR23b,D23,GMLMS24},
ideal gases~\cite{ZMHM22},
phase transitions~\cite{HR22},
active matter~\cite{SL22},
the autonomous information engine \cite{CBZH23},
ionic liquids~\cite{CWSKP24},
polymers~\cite{LLLHJ23,LLLHLJ23},
Langevin systems~\cite{BRP23,BR23}, and
plasmas~\cite{DDK23}. Particularly noteworthy is the recent surge in studies examining the effect in quantum systems~\cite{CLL21,CTH23,CTH23b,MAKP24,Joshietal24,SSMTARO24,WW24,CMA24,SPC24,YAC24,LZYZ24}.
The escalating interest in Mpemba-like effects is illustrated by Fig.~\ref{fig00}.

\begin{figure}
      \includegraphics[width=0.9\columnwidth]{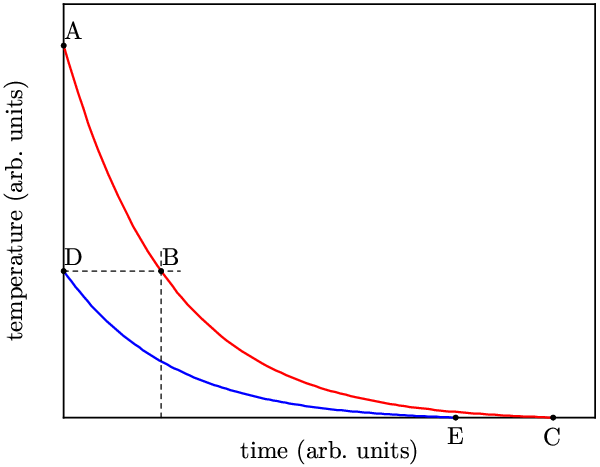}
      \caption{Sketch of supposed cooling curves for two
similar systems in the same environment, adapted from Ref.~\cite{O79}.
  \label{fig0}}
\end{figure}

To elucidate the counterintuitive nature of the Mpemba effect, consider the following excerpt from Osborne~\cite{O79}:
\begin{quotation}
\noindent
``Why is the effect unexpected? We suppose the rate
of heat transfer to depend on the temperature
difference between the cooling system and its
environment, and not to depend on its previous
history. This is represented in Fig.~\ref{fig0}, a sketch of
supposed cooling curves showing temperature against
time for two similar systems placed in the same
environment at the same time but at different initial
temperatures. The hot water takes a finite time to cool
down to the starting temperature of the cooler water.
We expect this system then to be identical to the
cooler system when it was first placed in the freezer,
that is for the hot starter and the cooler system to be
identical at the points represented by B and D on the
sketch. Subsequent cooling should also be identical, so
that the cooling curve BC should be similar to the
cooling curve DE.

But the subsequent cooling is not identical (the
overtaking effect would be represented by the curve
BC cutting the curve DE). Hence the states of the two
systems when represented by the points B and D are
not identical. Any explanation for the overtaking
effect should enable us to describe the difference
between the two systems as represented by the points
B and D in such a way that we would expect the
system starting hotter to cool faster even below this
temperature. What differences might there be?''
\end{quotation}

Osborne's depiction in his first ``common-sense'' paragraph is aptly illustrated by Newton's law of cooling \cite{Newton1701,B12,D12},
\beq
\label{Newton}
\dot{T}(t)=-\lambda\left[T(t)-\Tb\right],
\eeq
where $\Tb$ is the temperature of the thermal bath (or environment) and $\lambda$ is the coefficient of heat transfer (or cooling rate), here assumed to be a (positive) constant.
Its value  depends on various factors including the material properties, the object geometry, the surface conditions, and the surrounding environment. For instance, in the case of water, $\lambda$ can typically range from about $10^{-3}~\text{s}$ to $10^{-2}~\text{s}$ depending on the temperature difference and the volume of water~\cite{quickfield}.

The general solution to Eq.~\eqref{Newton} is simply
\beq
\label{2}
T(t)=\Tb+\left(T_0-\Tb\right)e^{-\lambda t},
\eeq
with $T_0=T(0)$ being the initial temperature.
On the other hand, if the Mpemba effect exists in a certain material, memory dynamics must be taken into account, so that ``the overtaking
effect should enable us to describe the difference
between the two systems as represented by the points
B and D [see Fig.~\ref{fig0}] in such a way that we would expect the
system starting hotter to cool faster even below this
temperature.''
The simplest way to incorporate those memory phenomena into Newton's law involves postulating  that the temperature's rate of change at time $t$ depends on the temperature at a preceding time $t-\tau$ \cite{HMMBE22}, i.e.,
\beq
\label{4}
\dot{T}(t)=-\lambda\left[T(t-\tau)-\Tb(t)\right],
\eeq
where $\tau>0$ is the delay time and we have considered the possibility that the bath temperature, $\Tb(t)$, changes with time.
Equation \eqref{4} can be considered as the simplest phenomenological equation incorporating memory effects.

Additionally to the Mpemba effect, another fascinating memory phenomenon is the Kovacs effect, originally reported in polymer materials \cite{K63,KAHR79}  and also observed in other complex systems \cite{MS04,PB10,PT14,TP14,RP14,KSI17,LGAR17,PP17b,LVPS19,SP21,MS22,PSPP23}. In the Kovacs effect, a sample, initially in equilibrium at a high temperature $\Tbm$, undergoes rapid quenching to a lower temperature $\Tbp$. The sample evolves for a specified waiting time $\tw$, but subsequently the bath temperature is abruptly raised to $T(\tw)$. Beyond $t=\tw$, the sample temperature exhibits a dynamic behavior: it initially increases, reaches a maximum, and then returns to equilibrium for a longer period. The effect thus highlights the nontrivial impact of the material's thermal past on its present and future behavior.

The aim of this paper is to analyze the Mpemba and Kovacs effects as described by the time-delayed cooling law, given by Eq.~\eqref{4}. Its general solution is studied in Sec.~\ref{sec2}, with special attention to single  and double quenches. It is seen that, in the former case,  the solution has a form similar to Eq.~\eqref{2}, except that the role of the exponential $e^{-\lambda t}$ is played by a function $\EE(t)$, here called the \tauE function. This function is positive definite only if the delay time is smaller than a threshold value, $\tau_{\max}=\lambda^{-1}e^{-1}$.
Next, in Sec.~\ref{sec3}, the solution is applied to the study of the Mpemba effect under simple protocols involving a hot bath at temperature $\Th$ and a cold bath at temperature $\Tc$. Sample A is thermalized at temperature $\Th$ and then quenched to temperature $\Tc$. Conversely, sample B is thermalized at $\Tc$, quenched to $\Th$, and, after a waiting time $\tw$, quenched again to the cold temperature $\Tc$. For simplicity, the quench of sample A and the first quench of sample B occur at the same time. After this preparation protocol, both samples are in the same environment (bath $\Tc$) but start with different temperatures. It is proved in Sec.~\ref{sec3} that (i) the existence or absence of the Mpemba effect is independent of the values of $\Th$ and $\Tc$, (ii) the Mpemba effect exists if and only if the two control parameters ($\tau$ and $\tw$) lie inside a certain narrow region, and (iii) the direct (cooling process) and the inverse (heating process) effects are fully equivalent. Section \ref{sec4} is devoted to the Kovacs effect and the function characterizing the strength of the Kovacs hump is identified. Interestingly, this hump points downward in the cooling process, thus signaling the presence of an \emph{anomalous} Kovacs effect \cite{PT14,LVPS19}, which becomes relatively stronger as both the delay time $\tau$ and the waiting time $\tw$ increase. Finally, the paper is closed in Sec.~\ref{sec5} with a summary and conclusion.

\section{Time-delayed Newton's cooling law}
\label{sec2}

\subsection{Expansion in powers of the delay time}
Before proceeding with the complete delayed equation, given by Eq.~\eqref{4}, let us consider $\tau$ as a small parameter and expand $T(t-\tau)$ in a series of powers of $\tau$:
\beq
\label{series}
T(t-\tau)=\sum_{k=0}^\infty \frac{d^k T(t)}{d t^k}\frac{(-\tau)^k}{k!}.
\eeq
Truncation at the level $k=0$ yields the original law, given by Eq.~\eqref{Newton}. Interestingly, truncation at $k=1$ just produces the same law, except for a renormalized cooling rate $\lambda'\equiv \lambda/(1-\lambda\tau)$.
However, the equation resulting from truncation at $k=2$ yields
\beq
\label{mu}
\mu\ddot{T}(t)+\dot{T}(t)=-\lambda'\left[T(t)-\Tb(t)\right],\quad \mu\equiv \frac{\lambda'\tau^2}{2}.
\eeq
The presence of the  term  $\mu\ddot{T}(t)$ suggests a more complex heat transfer process compared to the standard Newton's cooling equation. It represents  a sort of ``thermal inertia'' of the material, meaning that it might resist changes in temperature more than as described by Newton's law. In fact, by assuming $\Tb=\text{const}$, the general solution to Eq.~\eqref{mu} is
\bal
T(t)=&\Tb+\frac{T_0-\Tb}{\lambda_+'-\lambda_-'}\left(\lambda_+'e^{-\lambda_-' t}-\lambda_-'e^{-\lambda_+' t}\right)\nn
&+\frac{\dot{T}_0}{\lambda_+'-\lambda_-'}\left(e^{-\lambda_-' t}-e^{-\lambda_+' t}\right),
\eal
where
\beq
\lambda_{\pm}'\equiv\frac{1\pm\sqrt{1-4\lambda'\mu}}{2\mu},\quad \dot{T}_0\equiv\dot{T}(0).
\eeq
Thus, the temperature evolution is not solely determined by the initial temperature $T_0$, as is the case with Eq.~\eqref{2}, but also by the initial slope $\dot{T}_0$, thereby violating the simplistic depiction outlined in Fig.~\ref{fig0} and allowing memory effects to manifest.
It is worthwhile noting that the second-order differential equation, given by Eq.~\eqref{mu}, with $\Tb=\text{const}$ is equivalent to the following coupled set of two first-order differential equations:
\begin{subequations}
\beq
\dot{T}(t)=-\lambda'\left[T(t)-\Tb\right]+q(t),
\eeq
\beq
\dot{q}(t)=-{\lambda'}^2\left[T(t)-\Tb\right]-\left(\mu^{-1}-\lambda'\right)q(t).
\eeq
This two-variable scheme resembles the approach used to investigate the Mpemba and Kovacs effects in granular and molecular gases \cite{LVPS17,TLLVPS19,BPRR20,SP20,THS21,GKG21,BPR21,BPR22,MSP22,MS22b}.

\end{subequations}

Certainly, truncating Eq.~\eqref{series} at higher orders leads to differential equations of increasing order, thereby amplifying the influence of memory effects.

\subsection{General solution in Laplace space}
The general solution of Eq.~\eqref{4} depends not only on the initial temperature $T_0$ but it is actually a functional of the previous history $T(t)$ for $-\tau<t<0$.
Given the linear character of Eq.~\eqref{4}, it is convenient to define the Laplace transforms as
\beq
\label{5}
\wT(s)=\int_0^\infty dt\,e^{-s t}T(t),\quad \wTb(s)=\int_0^\infty dt\,e^{-s t}\Tb(t).
\eeq
Thus, Eq.~\eqref{4} becomes
\beq
\label{6}
s \wT(s)-T_0=\wTb(s)-e^{-s\tau}\wT(s)-\wH(s),
\eeq
where
\beq
\label{7}
\wH(s)=\int_{-\tau}^0 dt\,e^{-s(t+\tau)}T(t),
\eeq
and, for simplicity, we have taken $\lambda^{-1}=1$ as the unit of time.
Thus, the general solution in Laplace space is
\beq
\label{8}
\wT(s)=\frac{1}{s+e^{-s\tau}}\left[T_0+\wTb(s)-\wH(s)\right].
\eeq

\subsection{Basic solution. Single quench}
Let us first consider the simple bath temperature history
\beq
\label{8b}
\Tb(t)=\begin{cases}
\Tbm,& t<0,\\
\Tbp,& t>0.
\end{cases}
\eeq
The system is assumed to have equilibrated at $\Tbm$ for $t<0$, that is, $T(t\leq 0)=\Tbm$. Then, at $t=0$, the system is quenched to the bath temperature $\Tbp$, so that $\lim_{t\to\infty}T(t)=\Tbp$.

Our aim is to describe the transient stage from $T(t\leq 0)=\Tbm$ to $T(\infty)=\Tbp$.
In Laplace space, this is described by Eq.~\eqref{8} with $T_0=\Tbm$, $\wTb(s)=\Tbp s^{-1}$, and
\beq
\label{9}
\wH(s)=\Tbm s^{-1}\left(1-e^{-s\tau}\right).
\eeq
Thus, Eq.~\eqref{8} reduces to
\beq
\label{10}
\wT(s)=\Tbp s^{-1}+\left(\Tbm-\Tbp\right)\wEE(s),
\eeq
where
\beq
\label{11}
\wEE(s)=s^{-1}-\frac{s^{-2}}{1+s^{-1}e^{-s\tau}}.
\eeq
Note that $\wEE(s)$ is independent of both $\Tbm$ and $\Tbp$.
In real time, one has
\beq
\label{12}
T(t)=\begin{cases}
\Tbm,& t\leq 0,\\
\Tbp +\left(\Tbm-\Tbp\right)\EE(t),& t\geq 0,
\end{cases}
\eeq
with $\EE(t)$ being the inverse Laplace transform of $\wEE(s)$. In the limit of no delay ($\tau\to 0$), $\wEE(s)\to (1+s)^{-1}$, so that $\EE(t)\to e^{-t}$ and Eq.~\eqref{2} is recovered.
Because of that, henceforth the quasi-exponential function $\EE(t)$ will be referred to as the  \tauE function .

In general, if $\tau>0$, the function $\EE(t)$ characterizes the transient of temperature from $\Tbm$ to $\Tbp$.
To obtain $\EE(t)$, let us expand $(1+s^{-1}e^{-s\tau})^{-1}=\sum_{n=0}^\infty (-s)^{-n}e^{-ns\tau}$ and rewrite Eq.~\eqref{11} as
\beq
\label{13}
\wEE(s)=s^{-1}-\sum_{n=0}^\infty (-s)^{-(n+2)}e^{-ns\tau}.
\eeq
Therefore,
\bal
\label{14}
\EE(t)=&1+\sum_{n=0}^\infty \frac{(n\tau-t)^{n+1}}{(n+1)!}\Theta(t-n\tau)\nn
=&1+\sum_{n=0}^{\lfloor{t/\tau}\rfloor} \frac{(n\tau-t)^{n+1}}{(n+1)!},\quad t\geq 0,
\eal
where $\Theta(\cdot)$ is the Heaviside step function and $\lfloor\cdot\rfloor$ denotes the floor function.
More explicitly,
\beq
\EE(t)=\begin{cases}
  1-t,&0\leq t\leq \tau,\\
  1-t+\frac{(\tau-t)^2}{2},&\tau\leq t\leq 2\tau,\\
  1-t+\frac{(\tau-t)^2}{2}+\frac{(2\tau-t)^3}{3!},&2\tau\leq t\leq 3\tau,\\
  \cdots,&\cdots
\end{cases}
\eeq
Note that
\beq
\label{dotEE}
\dot{\EE}(t)=-\begin{cases}
1,&0\leq t\leq \tau,\\
\EE(t-\tau),&t\geq \tau.
\end{cases}
\eeq

\begin{figure}
      \includegraphics[width=0.9\columnwidth]{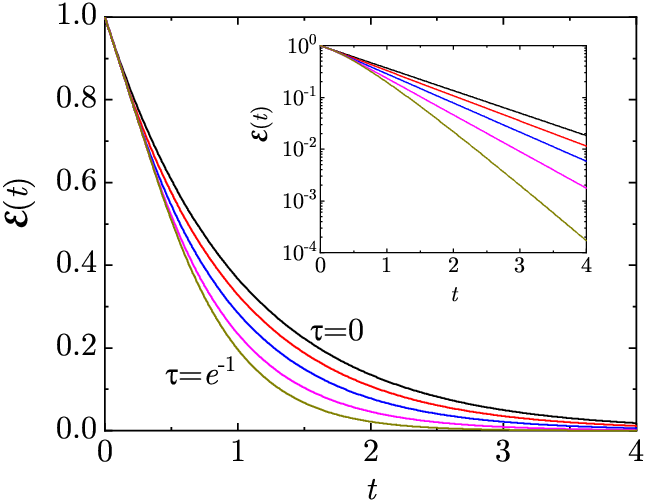}
      \caption{Plot of the \tauE function $\EE(t)$ with, from top to bottom, $\tau=0$, $0.1$, $0.2$, $0.3$, and $e^{-1}\simeq 0.368$. Inset; The graph in semilogarithmic scale.
  \label{fig1}}
\end{figure}

Figure~\ref{fig1} shows the \tauE function $\EE(t)$ for several values of $\tau$ in normal (main panel) and semilogarithmic (inset) scales. As we can see, the larger the delay time $\tau$, the faster the decay of $\EE(t)$.

The long-time behavior of $\EE(t)$ is governed by the dominant pole of $\wEE(s)$, that is, the root of the transcendental  equation $s+e^{-s\tau}=0$ with the least negative real part. It can be shown that such a root is real ($s=-\kappa$ with $\kappa<\tau^{-1}$) provided that $\tau<\tau_{\max}=e^{-1}\simeq 0.368$ (at which case $\kappa=e\simeq 2.718$). In that domain, $\kappa$ is the real root of $\kappa=e^{\kappa\tau}$, i.e., $\kappa=-\tau^{-1}W_0(-\tau)$, where $W_0(z)$ is the principal branch of the Lambert function.
If, on the other hand, $\tau>\tau_{\max}$, then the dominant root is a pair of complex conjugates, implying an oscillatory behavior with a negative absolute minimum $\EE_{\min}=\EE(t_{\min})<0$ at a certain time $t=t_{\min}$. The existence of this minimum  compromises the positive-definiteness of the solution, given by Eq.~\eqref{12}. Suppose that $\tau>\tau_{\max}=e^{-1}$ and $\Tbm/\Tbp>1+|\EE_{\min}|^{-1}$. Then, $T(t_{\min})/\Tbp=1-(\Tbm/\Tbp-1)|\EE_{\min}|<0$.
Therefore, the time-delayed Newton's cooling law is physically meaningful \emph{only if} $\tau<\tau_{\max}=e^{-1}$. This is the maximum value of $\tau$ considered in Fig.~\ref{fig1}.

\begin{figure}
      \includegraphics[width=0.9\columnwidth]{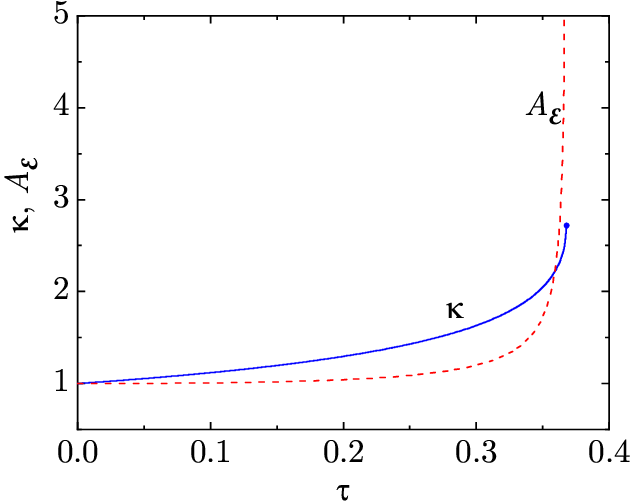}
      \caption{Plot of the damping coefficient $\kappa$ and the amplitude $A_{\EE}$ as  functions of $\tau$. The circle represents the damping coefficient $\kappa=e\simeq 2.718$ at the maximum delay time $\tau_{\max}=e^{-1}\simeq 0.368$.
  \label{fig2}}
\end{figure}

\begin{figure*}
\includegraphics[width=1.6\columnwidth]{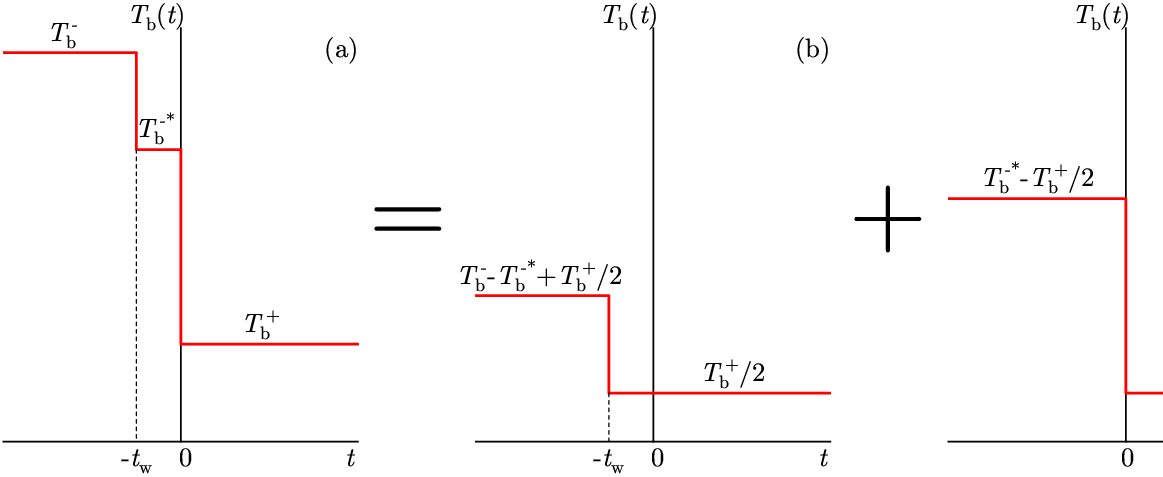}
      \caption{Schematic representation of the decomposition of the (a) double-quench protocol into the sum of the (b),(c) single-quench protocols.
  \label{fig3}
}
\end{figure*}

The amplitude $A_{\EE}$ of the asymptotic decay of $\EE(t)$ is given by the residue of $\wEE(s)$ at the pole $s=-\kappa$. Thus,
\beq
\label{AEE}
\EE(t\gg 1)\approx A_{\EE}e^{-\kappa t},\quad A_{\EE}=\frac{\kappa^{-1}}{1-\tau\kappa}.
\eeq
The damping coefficient $\kappa$ and the amplitude  $A_{\EE}$ are displayed as functions of the delay time $\tau$ in Fig.~\ref{fig2}. The damping coefficient increases from $\kappa=1$ at $\tau=0$ to $\kappa=e\simeq 2.718$ at the maximum delay time $\tau_{\max}=e^{-1}\simeq 0.368$. As for the amplitude, it is practically $A_{\EE}\simeq 1$ until $\tau\simeq 0.3$, but then it diverges as $\tau$ approaches its maximum value.

Suppose two samples (A and B) subjected to the quench described by Eq.~\eqref{8b}, except that the prior bath temperatures are different, say $\TbmA>\TbmB$. In that case, according to Eq.~\eqref{12},
\beq
\label{17}
\TA(t)-\TB(t)=\begin{cases}
\TbmA-\TbmB,& t\leq 0,\\
(\TbmA-\TbmB)\EE(t),& t\geq 0.
\end{cases}
\eeq
Since $\mathcal{E}(t)>0$ if $\tau<\tau_{\text{max}}$, Eq.~\eqref{17} proves that \emph{no} Mpemba effect is possible with the single-quench protocol.

\subsection{A more complex solution. Double quench}
Now, instead of taking the single-quench protocol, given by Eq.~\eqref{8b}, let us assume that the material is kept at a {prior} temperature $\Tp$ for times $t<-\tw$, next it is quenched to a {middle} bath temperature $\Tm$ for $-t_w<t<0$,  and then (after a waiting time $\tw$) it is quenched again to a {final} bath temperature $\Tbp$. This double-quench protocol is
\beq
\label{18}
\Tb(t)=\begin{cases}
\Tp,& t<-\tw,\\
\Tm,& -\tw<t<0,\\
\Tbp,&t>0.
\end{cases}
\eeq
A sketch of this protocol is shown in Fig.~\ref{fig3}(a). It can be expressed as the sum of the two single-quench protocols represented in Figs.~\ref{fig3}(b) and \ref{fig3}(c), respectively. Of course, the decomposition is still valid if an arbitrary constant is added and subtracted to each single-quench protocol, respectively. In general, an $n$fold-quench protocol is equivalent to the sum of $n$ single-quench protocols.

Taking into account the linear character of Eq.~\eqref{4}, the solution associated with the bath protocol given by Eq.~\eqref{18} can be obtained as the superposition of the solutions associated  with the protocols in Figs.~\ref{fig3}(b) and \ref{fig3}(c). Therefore,
\beq
T(t)=T_1(t)+T_2(t),
\eeq
where, by application of Eq.~\eqref{12},
\begin{subequations}
\beq
T_1(t)=\begin{cases}
\Tp-\Tm+\Tbp/2,&t\leq -\tw,\\
\Tbp/2+(\Tp-\Tm)\EE(t+\tw),&t\geq -\tw,
\end{cases}
\eeq
\end{subequations}
As a consequence, the solution corresponding to the double quench is
\beq
\label{24b}
T(t)=\begin{cases}
\Tp,&t\leq -\tw,\\
\Tm+(\Tp-\Tm)\EE(t+\tw),&-\tw\leq t\leq 0,\\
\Tbp+(\Tm-\Tbp)\EE(t)\\
+(\Tp-\Tm)\EE(t+\tw),&t\geq 0.
\end{cases}
\eeq
In particular, at $t=0$,
\beq
T_0=\Tm+(\Tp-\Tm)\EE(\tw).
\eeq

\begin{figure}
\includegraphics[width=0.45\columnwidth]{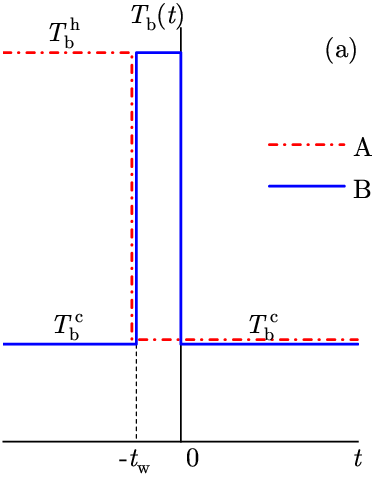}
\includegraphics[width=0.45\columnwidth]{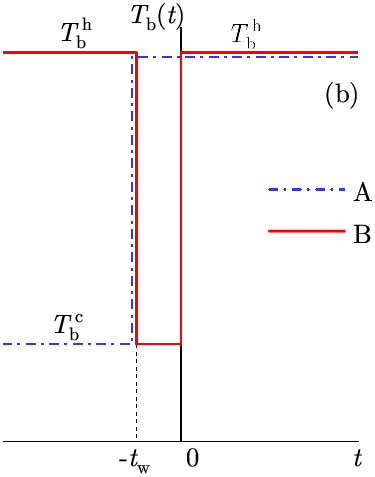}
      \caption{Schematic representation of the protocols A and B for (a) the direct Mpemba effect and (b) the inverse Mpemba effect.
  \label{fig4}}
\end{figure}

\section{Mpemba effect}
\label{sec3}

To explore whether the Mpemba effect can be observed from the solutions of the time-delayed Newton's cooling equation, given by Eq.~\eqref{4}, we consider two thermal reservoirs, i.e., a hot one (at temperature $\Th$) and a cold one (at temperature $\Tc$). Two simple protocols (A and B) are proposed for the Mpemba effect, both direct and inverse, as illustrated in Fig.~\ref{fig4}. Let us first describe the protocols for the direct effect; see Fig.~\ref{fig4}(a). According to protocol A, one sample (A) is first equilibrated at the hot bath temperature $\Th$ and then it is suddenly quenched to the cold bath temperature $\Tc$ at time $t=-\tw$. The other sample (B) is first equilibrated at the cold bath temperature $\Tc$, then quenched to the hot bath temperature $\Th$ at $t=-\tw$, and finally quenched to the cold bath at $t=0$. Thus, sample A experiences a single quench ($\Th\to\Tc$), while sample B is subjected to a double quench ($\Tc\to\Th\to\Tc$). As seen from Fig.~\ref{fig4}(b), protocols A and B for the inverse Mpemba effect are the same, except for the exchange $\Th\leftrightarrow\Tc$ (i.e., A: $\Tc\to\Th$; B: $\Th\to\Tc\to\Th$).

For the protocols depicted in Fig.~\ref{fig4}(a), the temporal evolution of sample A is given by Eq.~\eqref{24b} with $\Tp=\Tm=\Th$ and $\Tbp=\Tc$. Analogously, in the case of sample B, $\Tp=\Tbp=\Tc$ and $\Tm=\Th$. Therefore,
\begin{subequations}
\label{23AB}
\beq
\label{23A}
\TA(t)=\begin{cases}
\Th,&t\leq -\tw,\\
\Tc+(\Th-\Tc)\EE(t+\tw),&t\geq -\tw,
\end{cases}
\eeq
\beq
\label{23B}
\TB(t)=\begin{cases}
 \Tc,&  \hspace{-0.2cm}t\leq -\tw,\\
\Th-(\Th-\Tc)\EE(t+\tw),&\hspace{-0.7cm}-\tw\leq t\leq 0,\\
\Tc+(\Th-\Tc)\left[\EE(t)-\EE(t+\tw)\right],& t\geq 0.
\end{cases}
\eeq
\end{subequations}

\begin{figure}
\includegraphics[width=0.8\columnwidth]{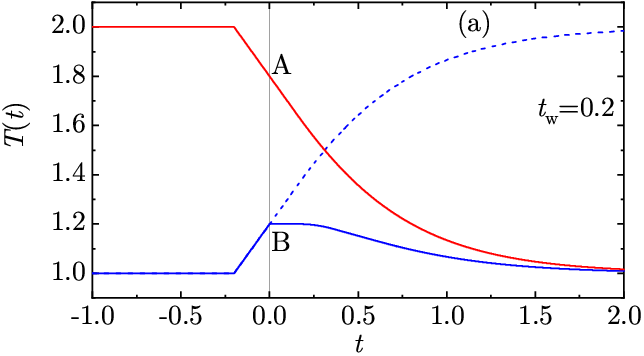}\\
\includegraphics[width=0.8\columnwidth]{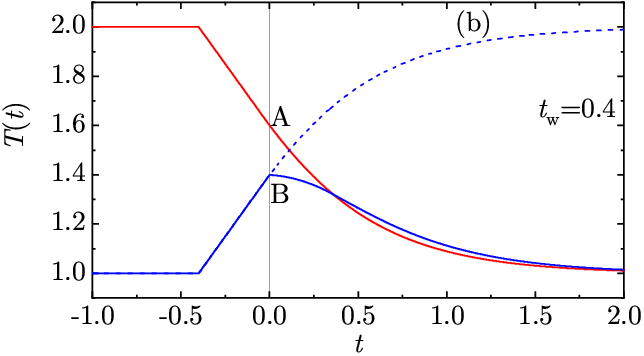}\\
\includegraphics[width=0.8\columnwidth]{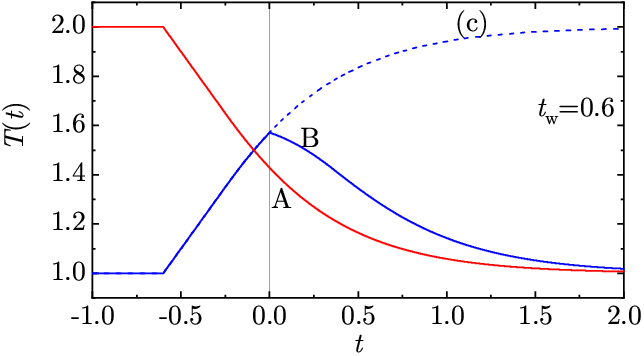}
      \caption{Plot of $\TA(t)$ and $\TB(t)$, including their history for $t<0$. Here,  $\Th=2$,  $\Tc=1$, and $\tau=0.36$, with (a) $\tw=0.2$, (b) $\tw=0.4$, and (c) $\tw=0.6$. In each panel, the dashed line depicts $\TB(t)$ if the sample were not quenched to $\Tc$ at $t=0$. Note that a Mpemba effect exists in the case of (b) ($\tw=0.4$).
        \label{fig5}}
\end{figure}

As an example, Fig.~\ref{fig5} shows the time evolution of $\TA(t)$ and $\TB(t)$ for a hot bath at $\Th=2$  (in arbitrary units), a cold bath at  $\Tc=1$, a delay time $\tau=0.36$, and three waiting times ($\tw=0.2,0.4,0.6$). As we can see, if the waiting time is too short (for instance, $\tw=0.2$), $\TB(0)$ is so far below $\TA(0)$ that no Mpemba crossing is possible. On the other hand, if the waiting time is too long (for instance, $\tw=0.6$), a crossing takes place during the preparation stage, that is, before sample B is quenched to $\Tc$ at $t=0$, so that $\TB(0)>\TA(0)$ and again no Mpemba effect occurs. However, if the waiting time is within the right range (for instance, $\tw=0.4$), one has $\TA(0)>\TB(0)$ but sample A cools down sufficiently faster than sample B as to eventually overtake it at a certain crossover time $\tx$ (direct Mpemba effect).

\begin{figure}
      \includegraphics[width=0.9\columnwidth]{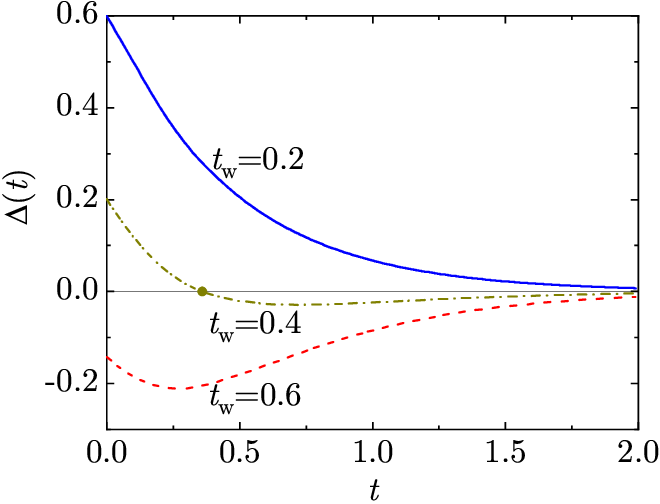}
      \caption{Plot of the difference function $\Delta(t)$ for $\tau=0.36$ and $\tw=0.2$ (solid curve), $0.4$ (dash-dotted curve), and $0.6$ (dashed curve).  The circle in the curve corresponding to $\tw=0.4$  denotes the corresponding crossover time $\tx$.
  \label{fig6}}
\end{figure}

Let us now investigate the necessary and sufficient conditions for the existence of the Mpemba effect.
From Eqs.~\eqref{23AB}, we have, for $t\geq 0$,
\beq
\TA(t)-\TB(t)=(\Th-\Tc)\Delta(t),\quad t\geq 0,
\eeq
where the difference function is
\beq
\label{Delta}
\Delta(t)=2\EE(t+\tw)-\EE(t).
\eeq
For the protocols shown in Fig.~\ref{fig4}(b), one would have $\TA(t)-\TB(t)=-(\Th-\Tc)\Delta(t)$. Thus, our first observation is that the existence or absence of the Mpemba effect (both direct and inverse) is independent of the bath temperatures $\Th$ and $\Tc$, depending only on the delay time $\tau$ and the waiting time $\tw$ through the difference function  $\Delta (t)$. Figure~\ref{fig6} shows the function $\Delta(t)$ for the same values of $\tau$ and $\tw$ as in Fig.~\ref{fig5}.

To prevent a situation similar to that illustrated in Fig.~\ref{fig5}(c), one must ensure that $\Delta(0)>0$, so that $\TA(0)-\TB(0)>0$ in the direct effect and $\TA(0)-\TB(0)<0$ in the inverse effect.  This implies that, for a given delay time $\tau$, the maximum value $\tw^{\max}$ of the waiting time $\tw$ is the solution of $\EE(\tw)=1/2$. This maximum value ranges from  $\tw^{\max}=\ln 2\simeq 0.693$ for $\tau=0$ to $\tw^{\max}\simeq 0.510$ for $\tau=e^{-1}$.

\begin{figure}
      \includegraphics[width=0.9\columnwidth]{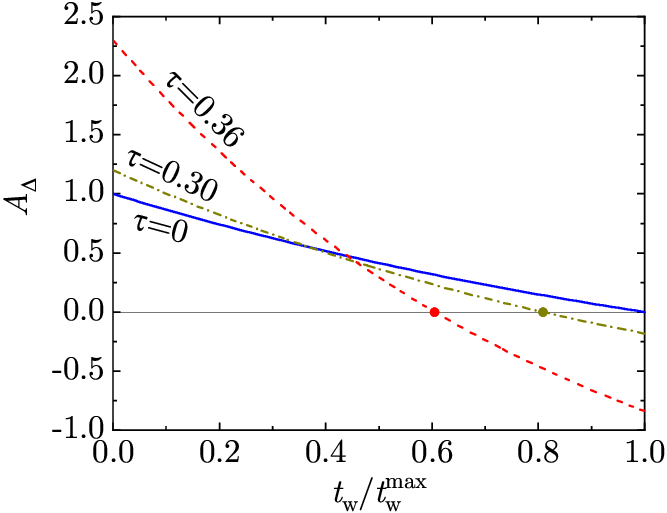}
      \caption{Plot of the amplitude $A_\Delta$ as a function of $\tw/\tw^{\max}$ for $\tau=0$ (solid curve), $0.30$ (dash-dotted curve), and $0.36$ (dashed curve). The circles in the curves corresponding to $\tau=0.30$ and $0.36$  denote the respective  values of $\twth/\tw^{\max}$.
  \label{fig7}}
\end{figure}

Next, once $\tw<\tw^{\max}$ and, therefore, $\Delta(t)>0$, the occurrence of a Mpemba effect requires that  $\Delta(\tx)=0$ at a certain crossover time $\tx$, so that it asymptotically relaxes to $\Delta(t)\to 0$ from below, as exemplified by the curve with $\tw=0.4$ in Fig.~\ref{fig6}. After the crossover time $\tx$, $-\Delta(t)$ presents a maximum at a time $\tM$ given by the solution to $\dot{\Delta}(t)=0$. According to Eqs.~\eqref{dotEE} and \eqref{Delta},
\beq
\dot{\Delta}(t)=-\begin{cases}
1,&0\leq t\leq \max\{0,\tau-\tw\},\\
2\EE(t+\tw-\tau)-1,&\max\{0,\tau-\tw\}\leq t\leq \tau,\\
\Delta(t-\tau),&t\geq \tau,
\end{cases}
\eeq
implying that $\tM=\tx+\tau$

From Eq.~\eqref{AEE}, we see that the asymptotic long-time behavior of $\Delta(t)$ is
\beq
\label{AD}
\Delta(t\gg 1)\approx A_{\Delta}e^{-\kappa t},\quad A_{\Delta}=\frac{\kappa^{-1}}{1-\tau\kappa}\left(2e^{-\kappa\tw}-1\right).
\eeq
The amplitude $A_\Delta$ is plotted in Fig.~\ref{fig7} as a function of $\tw/\tw^{\text{max}}$ for $\tau=0$, $0.3$, and $0.36$. Except in the undelayed case ($\tau=0$), we can see that $A_\Delta$ becomes negative if $\tw$ is larger than a threshold value $\twth$.
Therefore, given a delay time $\tau$, the minimum waiting time $\twth$ is determined by the condition $A_\Delta=0$, that is, $\twth=\kappa^{-1}\ln 2$.

\begin{figure}
      \includegraphics[width=0.9\columnwidth]{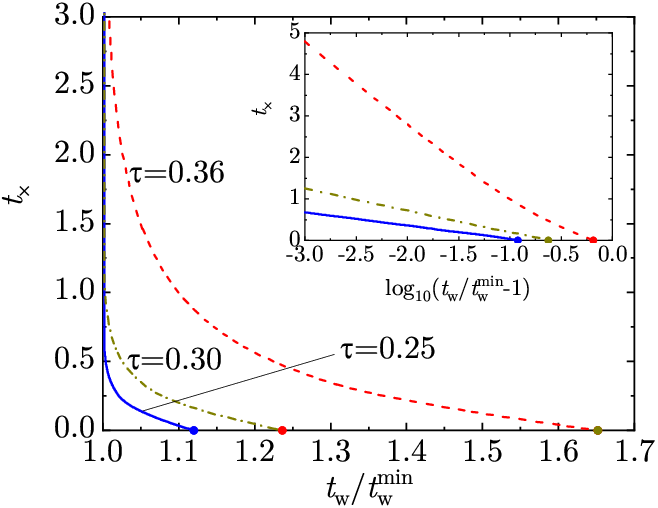}
      \caption{Plot of the crossover time $\tx$ as a function of $\tw/\twth$ for $\tau=0.25$ (solid curve), $0.30$ (dash-dotted curve), and $0.36$ (dashed curve). The circles   denote the respective  values of $\tw^{\max}/\twth$. Inset: $\tx$ vs $\log_{10}(\tw/\twth-1)$.
  \label{fig8}}
\end{figure}

\begin{figure}
      \includegraphics[width=0.9\columnwidth]{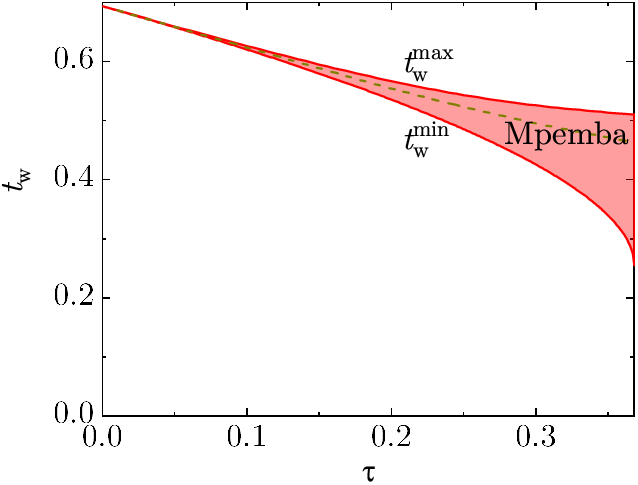}
      \caption{Phase space for the Mpemba effect. The lower and upper curves represent $\twth$ and $\tw^{\max}$, respectively, as functions of the delay time $\tau$. If $\tw<\twth$, then $\Delta(t)>0$ for $t\geq 0$.
      If, on the other hand, $\tw>\tw^{\max}$, then $\Delta(t)<0$ for $t\geq 0$. Therefore, the Mpemba effect occurs if and only if $\twth<\tw<\tw^{\max}$ (shaded region). The dashed line represents the locus defined by the condition $\Delta(0)=|\Delta(\tM)|$.
  \label{fig9}}
\end{figure}

\begin{figure*}
\includegraphics[width=0.7\columnwidth]{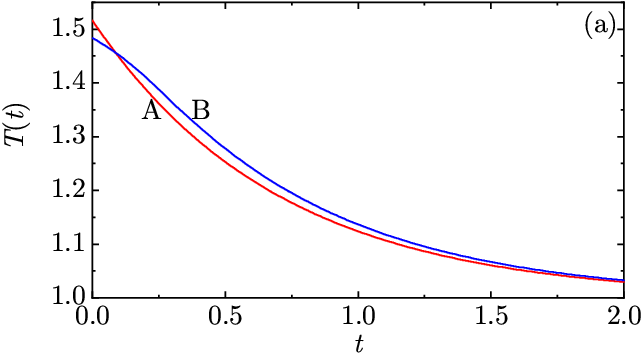}\includegraphics[width=0.7\columnwidth]{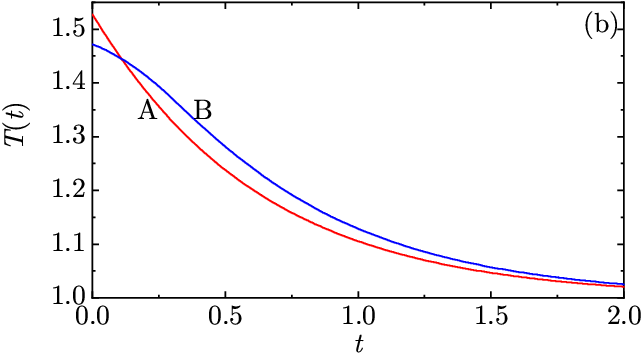}\includegraphics[width=0.7\columnwidth]{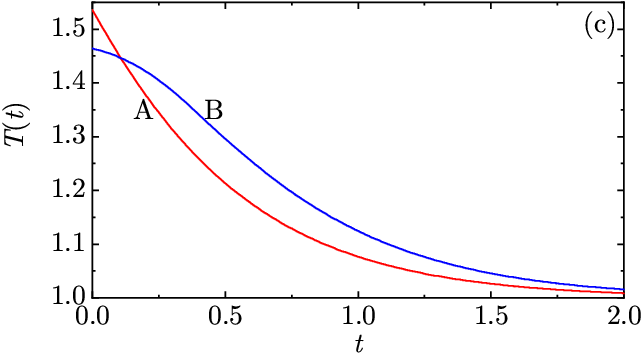}\\
\includegraphics[width=0.7\columnwidth]{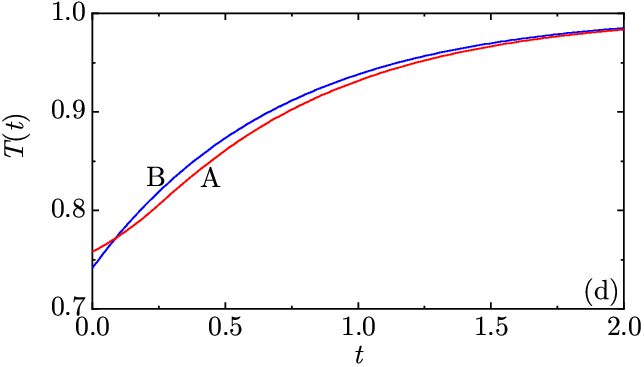}\includegraphics[width=0.7\columnwidth]{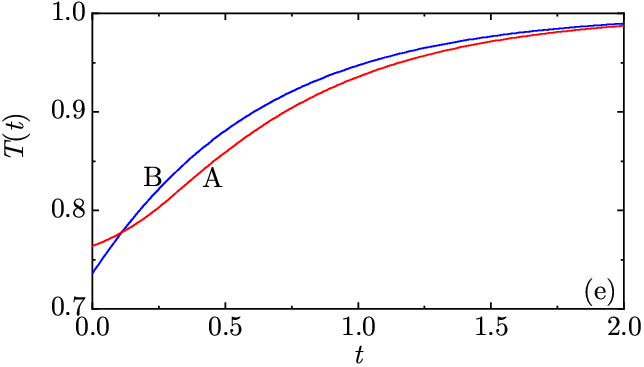}\includegraphics[width=0.7\columnwidth]{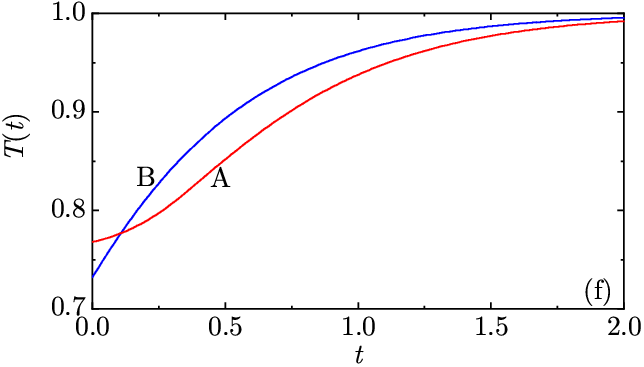}
      \caption{Plot of $\TA(t)$ and $\TB(t)$ for $t>0$. (a),(d) $(\tau,\tw)=(0.25, 0.52)$, (b),(e) $(\tau,\tw)=(0.30, 0.49)$, and (c),(f)  $(\tau,\tw)=(0.36, 0.47)$. (a)--(c) The direct Mpemba effect (with $\Tc=1$ and $\Th=2$); (d)--(f) the inverse Mpemba effect (with $\Tc=0.5$ and $\Th=1$).
  \label{fig10}}
\end{figure*}

If $\twth<\tw<\tw^{\text{max}}$, the crossover time $\tx$ is characterized by the condition $\Delta(\tx)=0$. It is plotted in Fig.~\ref{fig8} as a function of $\tw/\twth$ for $\tau=0.25$, $0.3$, and $0.36$. As $\tw$ increases from $\twth$ to $\tw^{\max}$, the crossover time $\tx$ decreases monotonically, vanishing at $\tw=\tw^{\max}$. Near $\tw=\twth$, $\tx$ diverges logarithmically as  $\tx\sim -\log_{10}(\tw/\twth-1)$ (see inset of Fig.~\ref{fig8}).

In summary, the time-delayed Newton's cooling equation exhibits a Mpemba effect (either direct or inverse) under the protocols depicted in Fig.~\ref{fig4} if, and only if, the two control parameters ($\tau$ and $\tw$) are such that the difference function $\Delta(t)$ fulfills two conditions: (i) $\Delta(0)>0$ (implying $\tw<\tw^{\max}$) and (ii) $\Delta(t)<0$ in the long-time regime (implying $\tw>\twth$).
The phase space for the occurrence of the Mpemba effect on the plane $\tw$ vs $\tau$ is shown in Fig.~\ref{fig9}.
Within the shaded Mpemba region, one can say that,  given a delay time $\tau$, the magnitude of the effect is maximal if the waiting time $\tw$ is such that the maximum positive value and the minimum negative value of $\Delta(t)$ are the same (except for the sign), i.e., $\Delta(0)=|\Delta(\tM)|$. That line of maximal Mpemba effect is also included in Fig.~\ref{fig9}.
To contextualize the characteristic values of the control parameters, note that, if $\lambda=10^{-3}\text{s}^{-1}$, then $\tau=0.36$ and $\tw=0.42$ correspond  to $6~\text{min}$ and $7~\text{min}$, respectively.

Some illustrations of the Mpemba effect (both direct and inverse) are displayed in Fig.~\ref{fig10}. The chosen values of $\tau$ and $\tw$ are close to the locus of maximal effect. We observe that, as expected, the Mpemba effect tends to become more pronounced as the delay time $\tau$ increases.

\begin{figure}
\includegraphics[width=0.45\columnwidth]{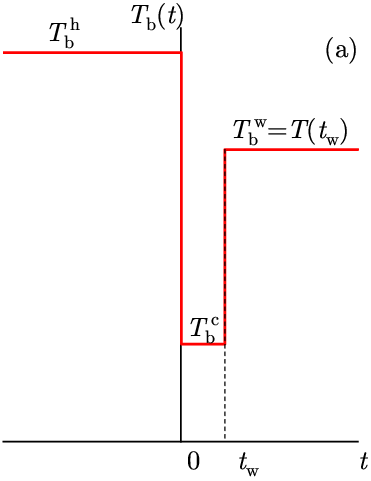}
\includegraphics[width=0.45\columnwidth]{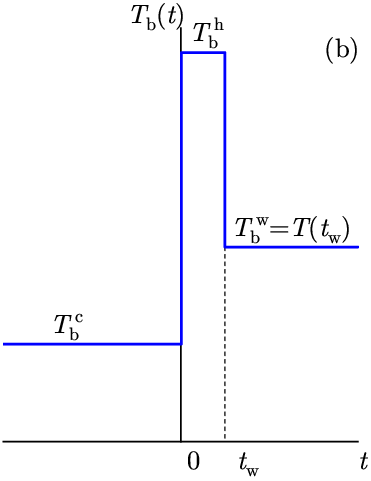}
      \caption{Schematic representation of the protocols  for (a) the direct Kovacs effect and (b) the inverse Kovacs effect.
  \label{fig11}}
\end{figure}

\begin{figure}
      \includegraphics[width=0.45\columnwidth]{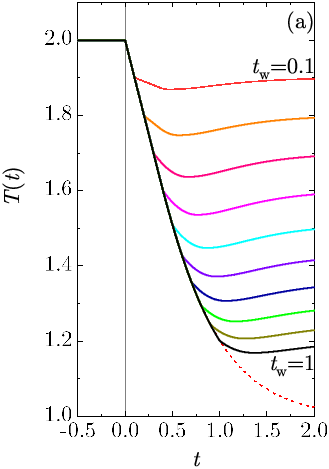}
      \includegraphics[width=0.45\columnwidth]{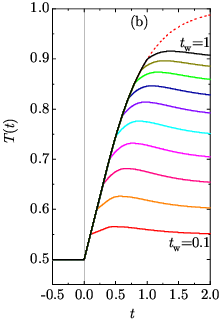}
      \caption{Kovacs effect for $\tau=0.36$ and $\tw=0.1$, $0.2$, \ldots, $1$. (a) The direct effect (with $\Th=2$ and $\Tc=1$); (b) the inverse effect (with $\Tc=0.5$ and $\Th=1$). In each panel, the dashed line represents $T(t)$ if the sample were not quenched to $T(\tw)$ at $t=\tw$.
  \label{fig12}}
\end{figure}

\begin{figure}
      \includegraphics[width=0.45\columnwidth]{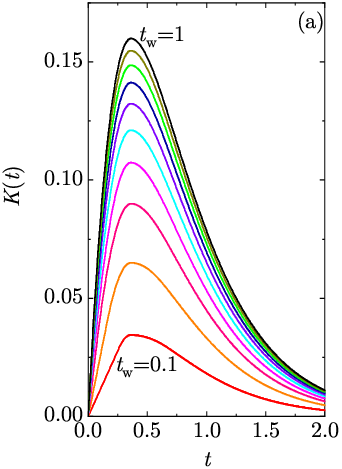}
      \includegraphics[width=0.45\columnwidth]{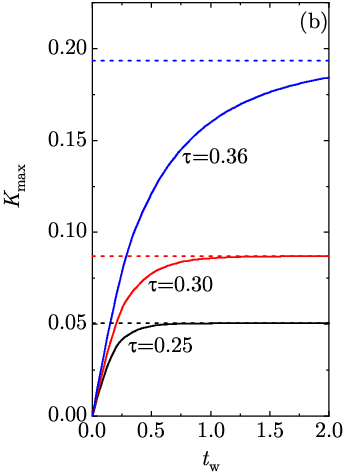}
      \caption{(a) Plot of the Kovacs hump function $\KK(t)$ for $\tau=0.36$ and $\tw=0.1$, $0.2$, \ldots, $1$. (b) Plot of the maximum value $\KK_{\max}$ of  $\KK(t)$ as a function of $\tw$ for, from bottom to top, $\tau=0.25$, $0.3$, and $0.36$. The dashed lines represent the asymptotic values $1-\tau-\kappa^{-1}$.
      \label{fig13}}
\end{figure}

\section{Kovacs effect}
\label{sec4}
In the Kovacs effect, the protocol is similar to that of sample B in the Mpemba effect (see Fig.~\ref{fig4}), except for a couple of points, as summarized in Fig.~\ref{fig11}. First, it is convenient to shift time by an amount $\tw$, so that the first quench occurs at $t=0$ and the second one at a waiting time $t=\tw$. Second, the final bath temperature is made to coincide with the instantaneous system's temperature at $t=\tw$, i.e., $\Tbp=T(\tw)\equiv \Tw$.
Thus, according to Eq.~\eqref{24b}, the temperature in the direct Kovacs effect is
\beq
\label{Kovacs}
T(t)=\begin{cases}
\Th,&t\leq 0,\\
\Tc+(\Th-\Tc)\EE(t),&0\leq t\leq \tw,\\
\Tw+(\Tc-\Tw)\EE(t-\tw)\\
+(\Th-\Tc)\EE(t),&t\geq \tw,
\end{cases}
\eeq
with
\beq
\label{Tw}
\Tw=\Tc+(\Th-\Tc)\EE(\tw).
\eeq
In the inverse effect, one must set $\Th\leftrightarrow \Tc$.

Figure \ref{fig12} illustrates the direct and inverse effects for $\tau=0.36$. As can be seen, after the quench at $t=\tw$, the temperature presents a local minimum (maximum) in the direct or cooling  (inverse or heating) case.
In the direct Kovacs effect, the temperature slope experiences a discontinuity from $\dot{T}(\tw^-)=-\left[T(\tw-\tau)-\Tc\right]$ to $\dot{T}(\tw^+)=-\left[T(\tw-\tau)-\Tw\right]$, but the sign is negative at both sides of $t=\tw$. Consequently, the Kovacs hump appears below $T=\Tw$, thus qualifying as an \emph{anomalous} Kovacs effect \cite{PT14,LVPS19}. An analogous conclusion holds in the inverse effect, where $\dot{T}(\tw^-)=\Th-T(\tw-\tau)>\dot{T}(\tw^+)=\Tw-T(\tw-\tau)>0$ and the hump appears above $T=\Tw$.

Let us characterize the Kovacs hump in more detail. From Eqs.~\eqref{Kovacs} and \eqref{Tw}, one finds that the temperature in the domain $t\geq \tw$ is
\begin{subequations}
\beq
T(t)-\Tw=-(\Tw-\Tc)\KK(t-\tw),\quad \text{(direct effect)},
\eeq
\beq
T(t)-\Tw=(\Th-\Tw)\KK(t-\tw),\quad \text{(inverse effect)},
\eeq
\end{subequations}
where
\beq
\KK(t)=\EE(t)-\frac{\EE(t+\tw)}{\EE(\tw)}
\eeq
is a semi-definite positive function, henceforth named the Kovacs hump function, which characterizes the relative strength of the Kovacs effect. It vanishes  both at $t=0$ and in the limit $t\to\infty$. Note that
\beq
\dot{\KK}(t)=-\begin{cases}
1-\frac{1}{\EE(\tw)},&0\leq t\leq \max\{0,\tau-\tw\},\\
1-\frac{\EE(t+\tw-\tau)}{\EE(\tw)},&\max\{0,\tau-\tw\}\leq t\leq \tau,\\
\KK(t-\tau),&t\geq \tau.
\end{cases}
\eeq
Therefore, $\dot{\KK}(\tau)=0$, implying that $\KK(t)$ has a maximum value $\KK_{\max}=1-\tau-\EE(\tau+\tw)/\EE(\tw)$  at a time $t_{\text{K}}=\tau$. At a given delay time $\tau$,  $\KK_{\max}$ increases monotonically with increasing waiting time $\tw$, reaching a finite value in the limit $\tw\to\infty$. From Eq.~\eqref{AEE}, we have $\lim_{\tw\to\infty}\KK_{\max}=1-\tau-e^{-\kappa\tau}=1-\tau-\kappa^{-1}$.

Figure \ref{fig13}(a) shows the Kovacs hump function  $\KK(t)$ for a delay time $\tau=0.36$ and several values of the waiting time. The maximum value $\KK_{\max}$ is plotted in Fig.~\ref{fig13}(b) as a function of $\tw$ for several values of $\tau$. As expected, the relative strength of the Kovacs effect, as measured by $\KK_{\max}$, increases with increasing $\tau$ and $\tw$.

\section{Conclusions}
\label{sec5}

\begin{table}
\caption{\label{table1}Summary of the main quantities and results.}
\begin{ruledtabular}
\begin{tabular}{ccc}
Quantity&Symbol&Expression\\
\hline
\multicolumn{3}{c}{General}\\
\hline
Delay time&$\tau$&Free parameter\\
Maximum delay time&$\tau_{\max}$&$e^{-1}\simeq 0.368$\\
\tauE function&$\EE(t)$&$\displaystyle{1+\sum_{n=0}^{\lfloor{t/\tau}\rfloor} \frac{(n\tau-t)^{n+1}}{(n+1)!}}$\\
Asymptotic decay of $\EE(t)$&&$\EE(t)\approx A_{\EE}e^{-\kappa t}$\\
Damping coefficient &$\kappa$&$\displaystyle{-\tau^{-1}W_0(-\tau)}$\\
Amplitude &$A_{\EE}$&$\displaystyle{\frac{\kappa^{-1}}{1-\tau\kappa}}$\\
Waiting time&$\tw$&Free parameter\\
\hline
\multicolumn{3}{c}{Mpemba effect}\\
\hline
Difference function&$\Delta(t)$&$2\EE(t+\tw)-\EE(t)$\\
Minimum waiting time&$\twth$&$\kappa^{-1}\ln 2$\\
Maximum waiting time&$\tw^{\max}$&Root of $\EE(\tw)=1/2$\\
Crossover time&$\tx$&Root of $\Delta(\tx)=0$\\
Time when $-\Delta(t)$ is maximum&$\tM$&$\tx+\tau$\\
\hline
\multicolumn{3}{c}{Kovacs effect}\\
\hline
Kovacs hump function&$\KK(t)$&$\displaystyle{\EE(t)-\frac{\EE(t+\tw)}{\EE(\tw)}}$\\
Time when $\KK(t)$ is maximum&$t_{\text{K}}$&$\tau$\\
Maximum value of $\KK(t)$&$\KK_{\max}$&$\displaystyle{1-\tau-\frac{\EE(\tau+\tw)}{\EE(\tw)}}$\\
\end{tabular}
\end{ruledtabular}
\end{table}

The time-delayed Newton's cooling law stands out as a seemingly straightforward yet powerful phenomenological model for grasping the intricacies of thermal memory dynamics. This paper has focused on unraveling its solution and applying it to two paradigmatic memory phenomena: the Mpemba and Kovacs effects.
The main quantities and results of this work are summarized in Table~\ref{table1}.

A pivotal role is played by the \tauE function $\EE(t)$, functioning as the delay-time analog of the conventional decaying exponential. Imposing the physical condition that $\EE(t)>0$ for all time sets an upper bound, $\tau<\tau_{\max}=e^{-1}$, for the delay time.

As the simplest protocol for the observation of the Mpemba effect, we have assumed that samples A and B were thermalized in the past ($t<-\tw$) to the temperatures ($\Th$ and $\Tc$) of a hot and a cold bath, respectively. Sudden quenches at $t=-\tw$ are then applied to both samples by exchanging their baths, followed by a second quench  to temperature $\Tc$ at $t=0$ for sample B.
Notably, the delay time enhances the cooling of sample A while inhibiting that of sample B, leading to the Mpemba effect under specific conditions ($\twth<\tw<\tw^{\max}$). The effect is independent of the bath temperatures ($\Th$ and $\Tc$) and is determined by a difference function $\Delta(t)$ parameterized by the control parameters $\tau$ and $\tw$.
In the inverse Mpemba effect, the protocol is identical, except for the exchange $\Th\leftrightarrow\Tc$, so that the difference function $\Delta(t)$ is the same as in the direct case. Therefore, the time-delayed Newton's cooling law with a \emph{constant coefficient of heat transfer} fails to capture the known fact that heating is faster than cooling \cite{LG20,VH21,IDLGR24}.

In general, the Mpemba effect necessitates the consideration of at least three distinct temperatures: the initial temperatures, $\TA(0)$ and $\TB(0)$, of both samples, and the temperature, $\Tc$, of the shared final bath. The protocol outlined in Fig.~\ref{fig4}(a) is relatively straightforward, involving only two bath temperatures ($\Th$ and $\Tc$), along with a waiting time ($\tw$). However, more complex protocols can be devised. For instance, sample B might initially equilibrate to a temperature $\TbmB$ (with $\Tc\leq\TbmB<\Th$) for $t<-\tw$, then be quenched at $t=-\tw$ to a bath at temperature $\TmB$ (with $\TbmB<\TmB\leq \Th$), and finally quenched at $t=0$ to the cold bath at temperature $\Tc$. Moreover, the introduction of two waiting times instead of just one could be contemplated, providing flexibility to accommodate different experimental protocols.

The exact solution of the time-delayed cooling equation under a double-quench protocol has also been exploited to study the Kovacs effect. While the system is relaxing from a hot bath temperature ($\Th$) to a cold bath temperature ($\Tc$), it is suddenly put in contact at a waiting time $\tw$ with a bath at temperature $\Tw=T(\tw)$. Instead of maintaining that temperature for $t>\tw$, the memory of a higher temperature in the past ($t<\tw$) makes the system momentarily keep cooling, eventually reaching a minimum temperature at time $t=\tw+\tau$, and finally relaxing to the bath temperature $\Tw$ from below. This downward hump contrasts with the  original one in polymer systems \cite{K63,KAHR79} but agrees with the anomalous Kovacs effect observed in granular gases \cite{PT14,LVPS19}.
In analogy with the Mpemba case, a relative strength measure, the Kovacs hump function $\KK(t)$,  remains independent of bath temperatures ($\Th$ and $\Tc$) and increases with $\tau$ and $\tw$.

It must be emphasized that. for simplicity, the analysis carried out  in this paper has assumed a constant coefficient of heat transfer $\lambda$. However,   factors such as dependence on the time-dependent temperature, variation in material properties with temperature, or changing boundary layer conditions due to natural convection might cause $\lambda$ to slightly deviate over time. While a constant $\lambda$ provides a valuable first approximation, acknowledging this potential time-dependence is important for a more comprehensive understanding of the Mpemba and Kovacs effect scenarios. Future investigations could explore the influence of a non-constant $\lambda$ on the results and potentially refine the model for greater accuracy.

In conclusion, the time-delayed Newton's law of cooling emerges  as a simplistic yet effective phenomenological model that accommodates memory effects. The exploration of the Mpemba and Kovacs effects within its framework  has unveiled significant insights into the intricate nature of these memory phenomena. This paper may contribute to a comprehensive analysis of the Mpemba and Kovacs effects, shedding light on their underlying mechanisms and expanding their relevance to diverse systems. The findings presented here not only can contribute to deepening our understanding of these memory phenomena, but also offer valuable insights applicable across various scientific domains, from physics to materials science. The exploration of time-delayed cooling laws paves the way for future research avenues, inviting further investigation into the fascinating interplay between thermal history, memory effects, and complex system behaviors.

\section*{Data availability}

The data that support the findings of this study are available from the author on reasonable request.

\acknowledgments

The author acknowledges financial support from Grant No.~PID2020-112936GB-I00 funded by MCIN/AEI/10.13039/501100011033, and from Grant No.~IB20079  funded by Junta de Extremadura (Spain) and by European Regional Development Fund (ERDF) ``A way of making Europe.''

\bibliography{C:/AA_D/Dropbox/Mis_Dropcumentos/bib_files/Granular}

\end{document}